\documentclass[conference,10pt,a4paper]{IEEEtran}
\usepackage{amsmath,amsthm,amssymb,bbm,enumerate}
\usepackage{graphicx,float,etoolbox}
\usepackage{xcolor,svg}
\usepackage{pgfplots,balance}
\usepackage{microtype}
\allowdisplaybreaks

\newtoggle{paper}
\togglefalse{paper}

\newcommand{\mc}{\mathcal}
\newcommand{\mrm}{\mathrm}

\newcommand{\norm}[1]{\left\| #1 \right\|}

\newcommand{\DSBS}{\mathsf{DSBS}}
\newcommand{\Rach}{\mc{R}_{\mrm{ach}}}
\newcommand{\RUL}{R^{\mrm{UL-SR}}_{\mrm{opt}}}
\newcommand{\RNO}{R^{\mrm{NO-SR}}_{\mrm{opt}}}

\title{Coordination Using Individually Shared Randomness}
\DeclareMathOperator*{\argmin}{arg\,min}
\theoremstyle{plain} \newtheorem{thm}{Theorem}
\theoremstyle{plain} 
\theoremstyle{plain} \newtheorem{defn}{Definition}
\theoremstyle{definition} \newtheorem{exmp}{Example}
\theoremstyle{plain} 
\theoremstyle{definition} 
\pgfplotsset{compat=1.14}
\usepackage[backend=biber, style=ieee]{biblatex}
\begin{document}
\author{%
  \IEEEauthorblockN{Gowtham~R.~Kurri, Vinod~M.~Prabhakaran}
  \IEEEauthorblockA{Tata Institute of Fundamental Research\\
                    Mumbai, India\\
                    Email: \{k.raghunath, vinodmp\}@tifr.res.in}
  \and
  \IEEEauthorblockN{Anand~D.~Sarwate}
  \IEEEauthorblockA{Rutgers, The State University of New Jersey\\
                    New Jersey, USA\\
                    Email:  anand.sarwate@rutgers.edu}
}

\maketitle

\begin{abstract}
Two processors output correlated sequences using the help of a coordinator with whom they individually share independent randomness. For the case of unlimited shared randomness, we characterize the rate of communication required from the coordinator to the processors over a broadcast link. We also give an achievable trade-off between the communication and shared randomness rates.  
\end{abstract}

\section{Introduction}

\begin{figure}[htbp]
\centering
\includegraphics[scale=1]{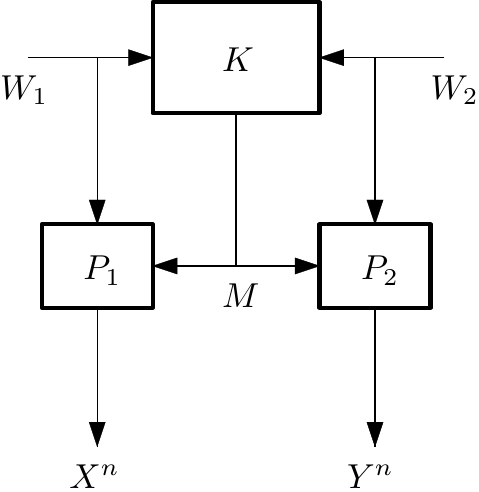}
\caption{Generating correlated random variables via shared randomness. $W_i$ is shared between coordinator $K$ and processor $P_i$, for $i=1,2$. $W_1$ is independent of $W_2$ and the goal is to produce $(X^n,Y^n)$ whose induced marginal distribution is close to i.i.d. sampling from a given $q(x,y)$. When $W_1,W_2$ are absent, this reduces to a model considered in~\cite{Wyner75}.}
\label{fig:model}
\end{figure}

We study the following problem, shown in Figure~\ref{fig:model}: two processors wish to output correlated random variables $X$ and $Y$, respectively\iftoggle{paper}{\footnote{\textit{A full version of this paper is accessible at: \url{http://www.tifr.res.in/~k.raghunath/ISIT2018/coordination.pdf}}}}{\ignorespaces}. In particular, they should output $n$ (approximately) independent and identically distributed (i.i.d.) samples $(X^n,Y^n)$ from a given joint distribution $q(x,y)$ of $(X,Y)$. They are assisted in this by a coordinator, who shares two independent rate-limited sources of randomness, one with each processor. The coordinator can send a rate limited common message to the processors to help them generate $(X^n,Y^n)$. Given a distribution $q(x,y)$ we are interested in the trade-off among the achievable rates.


Broadly, this problem belongs to a rich class of problems, widely studied in the literature, in which two or more agents connected through a network generate correlated random variables. An early work of this kind is due to Wyner~\cite{Wyner75} who characterized the minimum rate of common randomness required for two processors to produce (approximately) i.i.d. samples from a given joint distribution $q(x,y)$ of correlated random variables \iftoggle{paper}{$X,Y$;}{$X,Y$ (In the same paper, an alternative definition for common information in terms of a source coding problem~\cite{GrayW74} was also provided);} this rate is known as Wyner's common information of the random variables $X,Y$. Bennet et al.~\cite{Bennet02} and Cuff~\cite{Cuff13} studied a processor observing i.i.d. $X^n$ that sends a message to another processor to approximate a noisy channel $q_{Y|X}$ between them. In Cuff et al.~\cite{CuffPC10}, several two node and three node networks are studied, where the nodes try to produce correlated random variables.
The problem of generating correlated random variables via interactive communication has been studied by Gohari and Anantharm~\cite{GohariA11} and Yassaee et al.~\cite{YassaeeGA15}. Out previous work~\cite{PrabhakaranS13} studies a distributed sampling problem in which a genie can help two processors observing correlated sources to output correlated sequences.

In our problem, the coordinator has access to two independent random variables $W_1$ and $W_2$ each of which is shared with only one processor: we call these \emph{shared randomness} (SR). The processors want to generate sequences $X^n$ and $Y^n$ such that they are (approximately) distributed according to the desired distribution; we are interested in a notion which has been called strong coordination~\cite{CuffPC10}. We characterize the optimal transmission rate under \emph{unlimited shared randomness}. It is easy to infer from the literature~\cite{Wyner75,Bennet02,Cuff13} that a rate of $\min \left\{0.5C(X;Y), I(X;Y)\right\}$ is achievable under \emph{unlimited shared randomness}, where $C(X;Y):=\underset{X-U-Y}{\min}I(X,Y;U)$ is Wyner's common information~\cite{Wyner75}. To this end, on one hand, note that shared randomness can be converted to common randomness by the coordinator sending the $XOR$ of two individually shared random strings producing $2$ bits of common randomness for every bit sent. Then, Wyner's result~\cite{Wyner75} gives us that $0.5C(X;Y)$ is achievable. On the other hand, note that using their shared randomness, coordinator and processor $P_1$ can sample $X^n$ i.i.d. with distribution $q_X$. We can treat coordinator and processor $P_1$ as a single entity (encoder) having an input i.i.d. $X^n$, which sends a message $M$ to processor $P_2$ (decoder), which has to produce $Y^n$ according to the desired distribution, implying that $I(X;Y)$ is achievable~\cite{Bennet02,Cuff13}. Our new achievable scheme builds on these two ideas while strictly improving over $\min \left\{0.5C(X;Y), I(X;Y)\right\}$. In fact, our scheme gives a trade-off between the communication and shared randomness rates which turns out to be tight in two additional settings: 1) when the \emph{shared randomness rates approach zero}, as expected, the optimal transmission rate required turns out to be equal to Wyner's common information~\cite{Wyner75}, and 2) when $X$ and $Y$ are equal, we can completely characterize the rate region.

%
\section{Problem Definition}
\iftoggle{paper}
{%
\emph{Notation}: Four random variables $(X,Y,Z,W)$ are said to form a Markov chain denoted by $X-Y-Z-W$ if the probability distribution $p(x,y,z,w)$ satisfies $p(x,y,z,w)=p(y,z)p(x|y)p(w|z)$. %
We use a capital letter (e.g.\ $P_X$) to denote a random p.m.f.~(see other works~\cite{YassaeeAG14}, \cite{Cuff13}). %
For any two sequences of random p.m.f.'s $\{ P_{X^{(n)}} : n \in \mathbb{N} \}$ and $\{ Q_{X^{(n)}} : n \in \mathbb{N} \}$ on a sequence of sets $\{ \mathcal{X}^{(n)} : n \in \mathbb{N} \}$ (where $\mathcal{X}^{(n)}$ is arbitrary and can differ from the Cartesian product $\mathcal{X}^n$), we write $P_{X^{(n)}}\approx Q_{X^{(n)}}$ if $\lim_{n\rightarrow \infty}\mathbbm{E} \norm{ P_{X^{(n)}}-Q_{X^{(n)}} }_1=0$.}
{%
\emph{Notation}: All the random variables are taking values in finite sets. $p^{\text{Unif}}_{\mathcal{A}}$ is used to denote uniform probability distribution over the set $\mathcal{A}$. Four random variables $(X,Y,Z,W)$ are said to form a Markov chain denoted by $X-Y-Z-W$ if the probability distribution $p(x,y,z,w)$ satisfies $p(x,y,z,w)=p(y,z)p(x|y)p(w|z)$. Sequences $X_i,\dots,X_n$ are denoted by $X_i^n$, with $X^n:= X_1^n$. The total variation distance between two probability distributions $p(x)$ and $q(x)$ defined over a set $\mathcal{X}$ is defined by $\lVert p_X-q_X \rVert_1:=\sum_{x\in \mathcal{X}}\frac{1}{2}\lvert p(x)-q(x) \rvert$. We use capital letter to denote a random pmf (similar to \cite{YassaeeAG14}, \cite{Cuff13}), e.g.\ $P_X$. For any two sequences of random pmfs $\{ P_{X^{(n)}} : n \in \mathbb{N} \}$ and $\{ Q_{X^{(n)}} : n \in \mathbb{N} \}$ on a sequence of sets $\{ \mathcal{X}^{(n)} : n \in \mathbb{N} \}$ (where $\mathcal{X}^{(n)}$ is arbitrary and can differ from the Cartesian product $\mathcal{X}^n$), we write $P_{X^{(n)}}\approx Q_{X^{(n)}}$ if $\lim_{n\rightarrow \infty}\mathbbm{E} \norm{ P_{X^{(n)}}-Q_{X^{(n)}} }_1=0$. Similarly, we use $p_{X^{(n)}}\approx q_{X^{(n)}}$ for two non-random p.m.f.s to denote the total variation constraint $\lim_{n\rightarrow \infty}\lVert p_{X^{(n)}}-q_{X^{(n)}}\rVert_1=0$.}   

Our model consists of a coordinator $K$, two processors $P_1$ \& $P_2$ and a rate limited common communication link from the coordinator to the two processors. The coordinator has two independent sources of randomness each of which is shared with a different processor as shown in Figure~\ref{fig:model}. Here, $W_1,W_2$ are independent, uniform random variables distributed over alphabets $[1:2^{nR_1}]$ and $[1:2^{nR_2}]$ respectively. On observing $W_1$ and $W_2$, the coordinator $K$ produces a message $M\in[1:2^{nR}]$ according to a random map $p(m|w_1,w_2)$ and sends it over the common communication link to the two processors. Processor $P_1$ on observing $W_1$ and $M$ produces $X^n\in\mathcal{X}^n$ according to a random map $p(x^n|m,w_1)$. Similarly, processor $P_2$ produces $y^n\in\mathcal{Y}^n$ according to $p(y^n|m,w_2)$. The goal is to produce $X^n$ and $Y^n$ such that they are (approximately) distributed according to $q_{X,Y}^{(n)}(x^n,y^n):=\Pi_{i=1}^n q_{X,Y}(x_i,y_i)$.

\begin{defn}
An \emph{$(n,2^{nR},2^{nR_1},2^{nR_2})$ simulation code} consists of $\left(p(m|w_1,w_2),p(x^n|m,w_1),p(y^n|m,w_2)\right)$, where $m\in[1:2^{nR}]$, $w_1\in[1:2^{nR_1}]$ and $w_2\in[1:2^{nR_2}]$.
\end{defn}

The joint distribution of $(W_1,W_2,M,X^n,Y^n)$ and induced distribution $p^{\text{(induced)}}_{X^n,Y^n}$ on $(X^n,Y^n)$ are
\begingroup\makeatletter\def\f@size{9.5}\check@mathfonts
\def\maketag@@@#1{\hbox{\m@th\large\normalfont#1}}%
\begin{align*}
p( w_1,w_2, m, x^n, y^n ) &= \frac{ p(m|w_1,w_2)p(x^n|m,w_1)p(y^n|m,w_2) }{2^{n(R_1+R_2)}} \\
p^{(\text{induced})}(x^n,y^n)
&=
\sum_{w_1,w_2,m} p( w_1,w_2, m, x^n, y^n ).
\end{align*}\endgroup
\begin{defn}
A rate triplet $(R,R_1,R_2)$ is said to be \emph{achievable for} $q(x,y)$, if there exists a sequence of $(n,2^{nR},2^{nR_1},2^{nR_2})$ simulation codes such that 
\begin{equation}
\lim_{n\to\infty} \lVert p^{(\emph{induced})}_{X^n,Y^n}-q_{X,Y}^{(n)}\rVert_1=0.
\end{equation}
The \emph{simulation rate region} $\mathcal{R}$ is the closure of the set of all  achievable rate triplets $(R,R_1,R_2)$.
\end{defn}
\begin{defn}[Optimal transmission rate under unlimited shared randomness]
$\RUL$ is the infimum of all the rates $R$ such that there exist $R_1$ and $R_2$ so that $(R,R_1,R_2)\in\mathcal{R}$.
\end{defn}

\begin{defn}[Optimal transmission rate under no shared randomness]
$\RNO$ is the infimum of all the rates $R$ such that $(R,0,0)\in\mathcal{R}$. 
\end{defn}

\section{Results}
Let $\Rach$ be the set of all non-negative rate triplets $(R,R_1,R_2)$ such that
\begin{align} \label{eqn:1}
R+R_1&\geq I(X,Y;U,U_1),\nonumber\\
R+R_2&\geq I(X,Y;U,U_2),\nonumber\\
R&\geq I(U_1;U_2|U),\nonumber\\
R+R_1+R_2&\geq I(U_1;U_2|U)+I(X,Y;U,U_1,U_2),\nonumber\\
2R+R_1+R_2&\geq I(U_1;U_2|U)+I(X,Y;U)\nonumber\\
&\hspace{1cm}+I(X,Y;U,U_1,U_2),\nonumber\\
2R&\geq I(U_1;U_2|U)+I(X,Y;U).
\end{align}
for some p.m.f. $p(u,u_1,u_2|x,y)$ s.t $X-(U,U_1)-(U,U_2)-Y$.
\begin{thm}[Achievable region]\label{theorem:achievability}
$\Rach \subseteq \mc{R}$.
\end{thm}
We show this to be tight in some settings. When the shared randomness rates $R_1$ and $R_2$ are sufficiently large, we can characterize the optimal transmission rate.
\begin{thm}[Unlimited shared randomness]\label{theorem:unlimited_shared_randomness}
\begin{align}
&\RUL=\min\max\big\{I(X;Y|U),I(X,Y;U)\big\}\nonumber\\
&=\min \max \left\{I(X;Y|U), \frac{1}{2}\big(I(X,Y;U)+I(X;Y|U)\big)\right\}\nonumber,
\end{align}
where the minimum is over all conditional p.m.f.'s $p(u|x,y)$ with $|\mathcal{U}|\leq |\mathcal{X}||\mathcal{Y}|+2$.
\end{thm}
Let $C(X;Y)$ denote Wyner's common information \cite{Wyner75}, defined by
\begin{align}
C(X;Y):=\underset{{X-U-Y}}{\min}I(X,Y;U)\label{eqn:wyner_discussion}. 
\end{align}
As expected, when the shared randomness rates approach zero, the optimal transmission rate is equal to Wyner's common information, $C(X;Y)$ as stated in the following theorem.  
\begin{thm}[No shared randomness]\label{theorem:noshared}
$\RNO=C(X;Y).$
\end{thm}
For the case when $X$ and $Y$ are equal, we can completely characterize the simulation rate region as follows.
\begin{thm}\label{theorem:X=Y=Z}
Suppose the random variables $X = Y$ almost surely. Then the simulation rate region is given by set of all non-negative rate triplets $(R,R_1,R_2)$ such that
\begin{align*}
R+\min\left\{R_1,R_2\right\}&\geq H(X),\\\nonumber
R&\geq \frac{H(X)}{2}.
\end{align*}
\end{thm}

\section{Discussion} \label{section:discussion}

In this section, we discuss the intuition behind our achievability part by focusing on Theorem~\ref{theorem:unlimited_shared_randomness}. The discussion here will be informal. See Section~\ref{section:proofs} for precise details. As we discuss below, results from the literature~\cite{Wyner75,Bennet02,Cuff13} imply that when the shared randomness rates are large enough, a rate $R$ of $\min\left\{0.5C(X;Y),I(X;Y)\right\}$ is achievable, where $C(X;Y)$ is Wyner's common information. Our achievability scheme builds on the ideas behind this and we show by an example that our results strictly improve over $\min\left\{0.5C(X;Y),I(X;Y)\right\}$.

\emph{$0.5C(X;Y)$ is achievable:}  Consider $W_1,W_2$ each to be uniformly distributed on $[1:2^{nR}]$. We treat $W_1$ and $W_2$ each as an $nR$-length bit string. Let the coordinator transmit the bit string $M=W_1 \oplus W_2$ $(`\oplus$' denotes bit-wise $XOR)$ over common communication link to both the processors. Note that rate of transmission is $R$. From this both the processors can recover $(W_1,W_2)$ which is a common random variable uniformly distributed on $[1:2^{n(2R)}]$. Then, Wyner's result~\cite{Wyner75} shows that $2R\geq C(X;Y)$ is achievable, i.e., $R\geq 0.5C(X;Y)$ is achievable.

\emph{$I(X;Y)$ is achievable:} Using their shared randomness, the coordinator and processor $P_1$ sample $X^n$ i.i.d. with distribution $q_X$. Now, one can treat the coordinator and the processor $P_1$ together as a single entity (encoder) which on observing i.i.d. $X^n$ transmits a message $M$ of rate $R$ to the processor $P_2$ (decoder), which has to produce $Y^n$ to be distributed according to $q_{Y|X}^{(n)}$. Notice that both encoder and decoder have access to a sufficiently large amount of common randomness $W_2$. This is the channel simulation problem~\cite{Bennet02,Cuff13}, whose results imply that a rate $R$ of $I(X;Y)$ is achievable. 

\begin{figure}[htbp]
\centering
\includegraphics[scale=0.88]{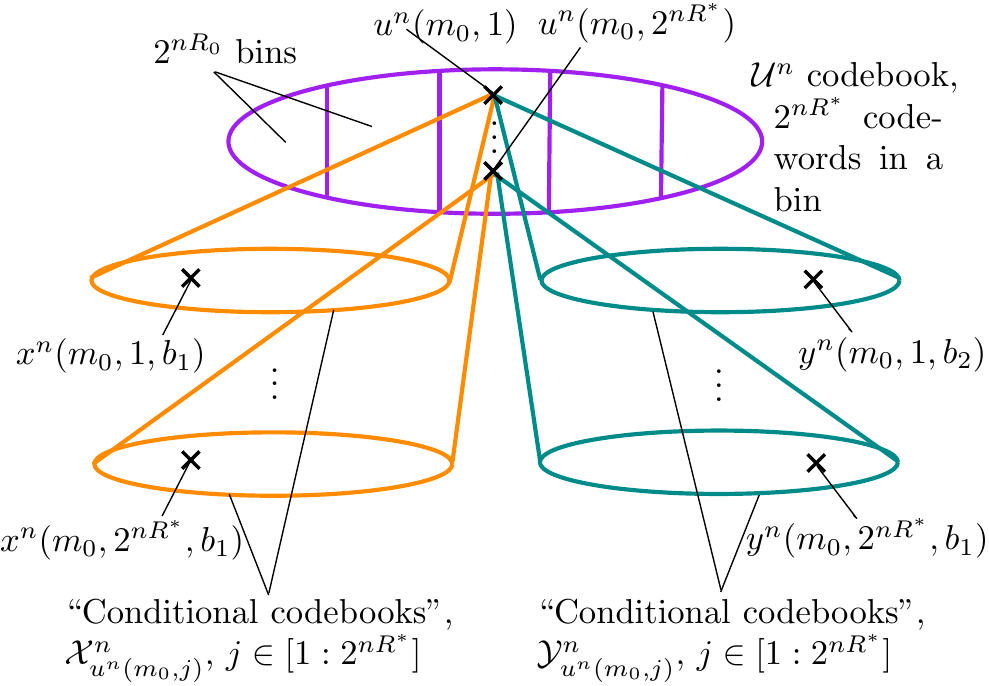}
\caption{A schematic diagram of coding scheme for the achievability part of Theorem~\ref{theorem:unlimited_shared_randomness} (when the shared randomness rates are large enough): Indices $m_0,b_1$ and $b_2$ are determined by the shared randomness in the following way: Index $m_0$ which is uniformly distributed on $[1:2^{nR_0}]$ is a concatenation of two $\frac{nR_0}{2}$ length bit strings $m_{01}$ and $m_{02}$, where $m_{0i}$ is obtained from shared randomness $w_i$, for $i=1,2$. Index $b_i$ which is independent of $m_0$ and uniformly distributed on $[1:2^{n\tilde{R}_i}]$ is also obtained from shared randomness $w_i$, for $i=1,2$. Note that $m_0,b_1,b_2$ are mutually independent of each other. The coordinator finds an $m^*$ inside the bin indexed by $m_0$, such that $\left(u^n(m_0,m^*),x^n(m_0,m^*,b_1),y^n(m_0,m^*,b_2)\right)$ is consistent with high probability. Loosely, $R^*>I(X;Y|U)$ ensures that there exists such an $m^*$. The coordinator then sends $(m_{01}\oplus m_{02},m^*)$ as a common message to the processors at a rate $R=\frac{R_0}{2}+R^*$. Note that $P_i$ has access to $m_{0i}$ and recovers $m_0$. The processors $P_1$ and $P_2$ output $x^n(m_0,m^*,b_1)$ and $y^n(m_0,m^*,b_2)$, respectively. Roughly, $R_0+R^*>I(X,Y;U)$ ensures that the output is according to the desired distribution. Since $R=\frac{R_0}{2}+R^*$, the above rate constraints imply that $\max \left\{I(X;Y|U),\frac{1}{2}\big(I(X;Y|U)+I(U;X,Y)\big)\right\}$ is achievable when the shared randomness rates are large enough.} 
\label{fig:picture}
\end{figure}
Our achievable scheme builds on these two ideas. While the complete technical details are in Section \ref{section:proofs}, an intuitive explanation is given in Figure~\ref{fig:picture}. 

It is easy to see that
\begin{align}
\RUL\leq\min\left\{0.5C(X;Y),I(X;Y)\right\}\label{eqn:discussion_lessthan_CInI}.
\end{align}
To see this, consider the second expression for $\RUL$ in Theorem~\ref{theorem:unlimited_shared_randomness}. Choosing $U$ to be a minimizer in \eqref{eqn:wyner_discussion} gives us $\RUL\leq 0.5C(X;Y)$. Choosing $U=\emptyset$ gives us $\RUL \leq I(X;Y)$. Next, we present an example where the inequality in \eqref{eqn:discussion_lessthan_CInI} is strict.

\begin{exmp}
Consider a doubly symmetric binary source $\DSBS(a)$ with joint distribution $q(x,y)=0.5(1-a)\delta_{xy}+0.5a(1-\delta_{xy})$, $a\in[0,0.5]$ and $x,y\in\{0,1\}$. For this distribution $I(X;Y)=1-h(a)$. Define $p^t(u|x,y) := tp^\bot(u|x,y)+(1-t)p^*(u|x,y), t\in[0,1],$ where
\begin{align*}
p^\bot(0|x,y)&=0.5=p^\bot(1|x,y)\ , \forall\  x,y,\\
p^*(u|x,y)&=\argmin_{p(u|x,y): X-U-Y}I(X,Y;U).
\end{align*}
The distribution $p^*(u|x,y)$ was found by Wyner~\cite{Wyner75}:
\begin{alignat*}{3}
p^*(0|0,1)&=p^*(1|1,0)  &&=0.5,\\
p^*(0|1,1)&=p^*(1|0,0) &&= 
b^2/(1-a),
\end{alignat*}
where $b=\frac{1}{2}\left(1- \sqrt{1-2a} \right)$ and the common information $C(X;Y) = I_{p^*}(X,Y;U)=1+h(a)-2h(b)$. Let $f(t)=\max\left\{I_{p^t}(X;Y|U),\frac{1}{2}\big(I_{p^t}(X,Y;U)+I_{p^t}(X;Y|U)\big)\right\}$, where $I_{p^t}(X;Y|U)$ and $I_{p^t}(X,Y;U)$ are calculated under $p^t(u|x,y)$:
\begin{align*}
&I_{p^t}(X,Y;U)=1+h(a)-h\left( \alpha,\frac{a}{2},\frac{a}{2},1-a-\alpha \right)\\
&I_{p^t}(X;Y|U)=2 h \left(\alpha+\frac{a}{2} \right)-h \left(\alpha,\frac{a}{2},\frac{a}{2},1-a-\alpha \right),
\end{align*}
where $\alpha=\left(1-t\right)b^2+\frac{t}{2}\left(1-a\right)$.
 It is easy to notice that $f(0)=0.5C(X;Y)$ and $f(1)=I(X;Y)$. We find a $t^*$ such that $I_{p^{t^*}}(X,Y;U)=I_{p^{t^*}}(X;Y|U)$, i.e., $t^*$ such that
 \begin{align*}
 1+h(a)&=2h\left(\left(1-t^*\right)b^2+\frac{t^*}{2}\left(1-a\right)+\frac{a}{2}\right)\\
 \Rightarrow \quad t^*&=\frac{1}{\left(\frac{1-a}{2}-b^2\right)}\left(h^{-1}\left(\frac{1+h(a)}{2}\right)-\frac{a}{2}-b^2\right).
 \end{align*}
For any $a\in(0,0.5)$, we can numerically see that $f(t^*)<\min\left\{f(0),f(1)\right\}=\min\left\{0.5C(X;Y),I(X;Y)\right\}$ (Figure~\ref{graphs} illustrates this fact for $a=0.1$ and $a=0.2$) implying that $\RUL<\min\left\{0.5C(X;Y),I(X;Y)\right\}$ since $\RUL \leq f(t^*)$. Moreover, we conjecture that $p^{t^*}(u|x,y)$ (with $t^*$ as identified above) is an optimizer for the expressions of $\RUL$ in Theorem~\ref{theorem:unlimited_shared_randomness}. The conjecture is supported by the fact that, it can be numerically checked that $p^{t^*}(u|x,y)$ is a minimizer among all the conditional p.m.f.'s $p(u|x,y)$ with $|\mathcal{U}|=2$.
\begin{figure}[htbp]
\centering
\newlength\fheight 
    \newlength\fwidth 
    \setlength\fheight{4.5cm} 
    \setlength\fwidth{7.6cm}
%
%
\definecolor{mycolor1}{rgb}{0.00000,0.44700,0.74100}%
\definecolor{mycolor2}{rgb}{1.00000,0.00000,1.00000}%
\begin{tikzpicture}

\begin{axis}[%
width=0.951\fwidth,
height=\fheight,
at={(0\fwidth,0\fheight)},
scale only axis,
xmin=0,
xmax=1,
xlabel style={font=\color{white!15!black}},
xlabel={$t$},
ymin=0.15,
ymax=0.55,
ylabel style={font=\color{white!15!black}},
ylabel={$f(t)$},
axis background/.style={fill=white},
legend style={at={(0.97,0.5)}, anchor=east, legend cell align=left, align=left, draw=white!15!black}
]
\addplot [color=mycolor1, line width=1.0pt]
  table[row sep=crcr]{%
0	0.436380283400076\\
0.00279999999999991	0.431499055555295\\
0.00600000000000001	0.426557884692081\\
0.00950000000000006	0.421713981411722\\
0.0134000000000001	0.4168360930299\\
0.0177	0.411951459591998\\
0.0224	0.407082762497141\\
0.0275000000000001	0.40224869478021\\
0.0329999999999999	0.39746451880138\\
0.0388999999999999	0.392742573837561\\
0.0452999999999999	0.388021828003421\\
0.0521	0.383391652121826\\
0.0593999999999999	0.37879591957208\\
0.0671999999999999	0.374251203473264\\
0.0756000000000001	0.369718728770988\\
0.0845	0.365268373707793\\
0.0940000000000001	0.360862568888059\\
0.1042	0.356474776169439\\
0.1151	0.352125858796536\\
0.1267	0.347832816323742\\
0.1391	0.343576307207822\\
0.1523	0.339374487447541\\
0.1663	0.33524216255484\\
0.1813	0.331138614803407\\
0.1972	0.327109445067336\\
0.2142	0.323121673675278\\
0.2323	0.319195303007056\\
0.2515	0.315347013931011\\
0.2719	0.311572886227332\\
0.2936	0.307872200330448\\
0.3167	0.304246959760662\\
0.3412	0.300715313268914\\
0.3436	0.300527573378146\\
0.3608	0.312924083055196\\
0.378	0.32494217372874\\
0.3952	0.336585280718171\\
0.4124	0.347856966449423\\
0.4296	0.358760853917631\\
0.4468	0.369300576421957\\
0.464	0.379479739618213\\
0.4812	0.389301892950686\\
0.4984	0.398770508260472\\
0.5156	0.407888963906333\\
0.5328	0.416660533132085\\
0.55	0.425088375711256\\
0.5672	0.433175532122551\\
0.5844	0.440924919678386\\
0.6016	0.448339330157219\\
0.6188	0.45542142858892\\
0.636	0.462173752918356\\
0.6532	0.468598714331232\\
0.6705	0.474733116308751\\
0.6878	0.480540847546466\\
0.7051	0.486023979838117\\
0.7224	0.491184460397482\\
0.7397	0.496024113604635\\
0.757	0.500544642801195\\
0.7743	0.504747632091284\\
0.7916	0.508634548114012\\
0.8089	0.512206741760539\\
0.8262	0.515465449814522\\
0.8435	0.518411796499324\\
0.8608	0.521046794918987\\
0.8781	0.523371348382871\\
0.8954	0.525386251606146\\
0.9127	0.5270921917801\\
0.93	0.528489749507709\\
0.9473	0.529579399600975\\
0.9646	0.530361511737473\\
0.9819	0.530836350974203\\
1	0.531004406410719\\
};
\addlegendentry{$a=0.1$}

\addplot [color=mycolor2, line width=1.0pt]
  table[row sep=crcr]{%
0	0.352952450491633\\
0.00570000000000004	0.346612136545404\\
0.0119	0.340182625832955\\
0.0185	0.333790024675116\\
0.0255000000000001	0.327444257460282\\
0.0328999999999999	0.321154344760102\\
0.0407	0.314928388317052\\
0.0489999999999999	0.308700802941808\\
0.0577000000000001	0.302559241628198\\
0.0669	0.29644380631286\\
0.0766	0.290369631440877\\
0.0868	0.284349818431383\\
0.0974999999999999	0.278395695856848\\
0.1087	0.272517046729866\\
0.1205	0.26667419966715\\
0.1328	0.26092792792823\\
0.1457	0.255241148140532\\
0.1592	0.249625991853822\\
0.1734	0.244054936128429\\
0.1882	0.23857965686195\\
0.2037	0.233173980752178\\
0.2199	0.227850460366026\\
0.2369	0.222590150843885\\
0.2546	0.217436417826332\\
0.2731	0.212371111851642\\
0.2924	0.207406515164156\\
0.3125	0.202553633843804\\
0.3335	0.19780045854464\\
0.3554	0.193160168361872\\
0.3782	0.188644769263606\\
0.4019	0.184265264727702\\
0.4265	0.180031818015786\\
0.4427	0.177497550666439\\
0.4659	0.185639679534829\\
0.4888	0.193353744760011\\
0.5116	0.2007111749631\\
0.5342	0.207682616998486\\
0.5567	0.214302586315447\\
0.5791	0.220573208483627\\
0.6014	0.226497077596756\\
0.6236	0.232077161518353\\
0.6457	0.237316725103683\\
0.6677	0.24221926776353\\
0.6897	0.246808623997168\\
0.7116	0.251065634178802\\
0.7335	0.255011557787713\\
0.7554	0.258646179287083\\
0.7772	0.261954891546168\\
0.799	0.264954916931875\\
0.8208	0.267646216875044\\
0.8426	0.270028784279182\\
0.8644	0.272102633732726\\
0.8862	0.273867793345498\\
0.9079	0.275318321415127\\
0.9296	0.276463055436915\\
0.9513	0.277302028131565\\
0.973	0.277835265984609\\
1	0.278071905112638\\
};
\addlegendentry{$a=0.2$}

\end{axis}
\end{tikzpicture}%
\caption{In both the plots $(t^*,f(t^*))$ is the minimum point which illustrates that $f(t^*)<\min\{f(0),f(1)\}$, where $(0,f(0))$ and $(1,f(1))$ are the respective corner points. (Top) Case when $f(0)=0.5C(X;Y)<I(X;Y)=f(1)$. $t^*=0.343436$ for $a=0.1$. (Bottom) Case when $f(0)=0.5C(X;Y)>I(X;Y)=f(1)$. $t^*=0.442523$ for $a=0.2$.}
\label{graphs}
\end{figure}
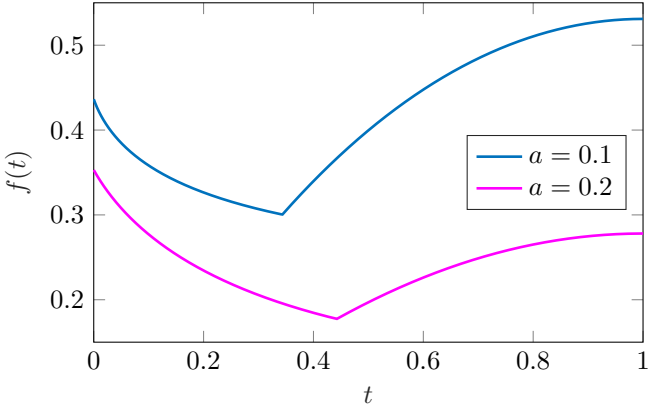
\end{exmp}
\section{Proofs} \label{section:proofs}
\begin{proof}[\textbf{Proof of Theorem~\ref{theorem:achievability}}]
The essential ideas of the proof were presented in Section~\ref{section:discussion} at an intuitive level. There, shared randomness rates were assumed to be large enough. Here we give a proof 
which formalizes the intuition and gives a trade-off among all the rates (i.e., without any assumption on the shared randomness rates). The proof is based on the OSRB (Output Statistics of Random Binning) framework developed by Yassaee, Aref and Gohari in~\cite{YassaeeAG14}. Let $(U^n,U_1^n,U_2^n,X^n,Y^n)$ be i.i.d. with distribution $p(u,u_1,u_2,x,y)$ such that $X-(U,U_1)-(U,U_2)-Y$ is a Markov chain. Bin indices $m_0,f,m^*,f_1,b_1,f_2,b_2$ with respective rates $R_0,\hat{R}_0,R^*,\hat{R}_1,\tilde{R}_1,\hat{R}_2,\tilde{R}_2$ are created from $(U^n,U_1^n,U^n_2)$ in a way that can be understood from the following joint probability distribution:
\begin{align}
P(u^n,&u_1^n,u_2^n,x^n,y^n,m_0,f,m^*,f_1,b_1,f_2,b_2)\nonumber\\
&\hspace{-1cm}=p(u^n,u_1^n,u_2^n)p(x^n|u^n,u_1^n)p(y^n|u^n,u_2^n)P(m_0,f|u^n)\nonumber\\
&\times P(m^*|u^n)P(f_1,b_1|u^n,u_1^n)P(f_2,b_2|u^n,u_2^n)\nonumber\\
&\hspace{-1cm}=P(b_1,b_2,f_1,f_2,m_0,f)P(u^n,u_1^n,u_2^n|b_1,b_2,f_1,f_2,m_0,f)\nonumber\\
&\times P(m^*|u^n)p(x^n|u^n,u_1^n)p(y^n|u^n,u_2^n)\label{eqn:ach_shrt_proof}.
\end{align}
Further, we use Slepian-Wolf decoders to estimate $(u^n,u_i^n)$ from $f_i,b_i,m_0,f,m^*$ for $i=1,2$.

Now, we impose a series of constraints on the rates 
 {\color{black}{and explain them}}. 
{\color{black}
\begin{align}
  R_0+\hat{R_0}& < H(U)\nonumber\\
  \tilde{R_1}+\hat{R_1}+R_0+\hat{R_0}& < H(U,U_1)\nonumber\\
  \tilde{R_2}+\hat{R_2}+R_0+\hat{R_0}& < H(U,U_2)\nonumber\\
  \tilde{R_1}+\hat{R_1}+\tilde{R_2}+\hat{R_2}+R_0+\hat{R_0}& < H(U,U_1,U_2)\label{eqn:theorem:arxiv_achievability_gather_1}
  \end{align}
  \begin{align}
 \tilde{R_1}+\hat{R_1}& > H(U_1|U)\nonumber\\
 \tilde{R_1}+\hat{R_1}+R_0+\hat{R_0}+R^*& > H(U,U_1)\nonumber\\
 \tilde{R_2}+\hat{R_2}& > H(U_2|U)\nonumber\\
 \tilde{R_2}+\hat{R_2}+R_0+\hat{R_0}+R^*& > H(U,U_2)\label{eqn:theorem:arxiv_achievability_gather_2}
\end{align}
}
The first set of constraints \eqref{eqn:theorem:arxiv_achievability_gather_1} ensure that $b_1,b_2,f_1,f_2,m_0,f$ are approximately (i.e., with vanishing total variation distance) uniformly distributed and mutually independent of each other~\cite[Theorem 1]{YassaeeAG14}. The second set of constraints \eqref{eqn:theorem:arxiv_achievability_gather_2} guarantees the success of Slepian-Wolf decoders with high probability~\cite[Lemma 1]{YassaeeAG14}. Thus, under these two sets of rate constraints \eqref{eqn:theorem:arxiv_achievability_gather_1} and \eqref{eqn:theorem:arxiv_achievability_gather_2}, the random p.m.f. comprising \eqref{eqn:ach_shrt_proof} and Slepian-Wolf decoders becomes approximately close to the p.m.f. below.
\begin{align}
\hat{P}(u^n,&u_1^n,u_2^n,x^n,y^n,m_0,f,m^*,f_1,b_1,f_2,b_2,\nonumber\\
&\hat{u}_1^n,\hat{u}_2^n,\hat{u}^n_{(1)},\hat{u}^n_{(2)})\nonumber\\
&\hspace{-1cm}=p^{\text{Unif}}(b_1)p^{\text{Unif}}(f_1)p^{\text{Unif}}(b_2)p^{\text{Unif}}(f_2)p^{\text{Unif}}(m_0)p^{\text{Unif}}(f)\nonumber\\
&\times P(u^n,u_1^n,u_2^n|b_1,b_2,f_1,f_2,m_0,f)P(m^*|u^n)\nonumber\\
&\times P^{SW}(\hat{u}^n_{(1)},\hat{u}_1^n|f_1,b_1,m_0,f,m^*)p(x^n|\hat{u}^n_{(1)},\hat{u}_1^n)\nonumber\\
&\times P^{SW}(\hat{u}^n_{(2)},\hat{u}_2^n|f_2,b_2,m_0,f,m^*)p(y^n|\hat{u}^n_{(2)},\hat{u}_2^n)\label{eqn:Thm_ach_protB}
\end{align}
The above p.m.f. is related to, but not exactly the same as that of our original problem due to the presence of conditional random p.m.f.'s and extra shared randomness. The connection between this p.m.f. and the original problem is described below. 
 In p.m.f. \eqref{eqn:Thm_ach_protB} we generate $b_1,b_2,f_1,f_2,m_0,f$ independently and uniformly from the  respective alphabets. We treat `$m_0$' as an $nR$-length string of bits i.e., a concatenation of two messages $m_{01},m_{02}$, each consisting of $\frac{nR_0}{2}$ bits. For $i=1,2$, we treat $m_{0i}$ and $b_i$ together as the shared randomness $w_i$ that is shared between coordinator and processor $P_i$. In addition, we have extra shared randomness $f,f_1$ and $f_2$ (to be eliminated later), where $f$ is shared among coordinator and both the processors and $f_i$ is shared between coordinator and processor $P_i$, for $i=1,2$.
 The coordinator on observing $b_1,b_2,f_1,f_2,m_0,f$ produces $u^n,u_1^n,u_2^n$ according to random p.m.f. $P(u^n,u_1^n,u_2^n|b_1,b_2,f_1,f_2,m_0,f)$ of \eqref{eqn:ach_shrt_proof} and sends $(m_{01}\oplus m_{02},m^*(u^n))$ as a common message $m$ to both the processors, where $m^*(u^n)$ is produced according to $P(m^*|u^n)$ of \eqref{eqn:ach_shrt_proof}. Thus, both the processors can recover `$m_0$' exactly since $P_i$ already has access to $m_{0i}$, for $i=1,2$. Then processor $P_i$ uses (random) Slepian-Wolf decoder $P^{SW}(\hat{u}^n_{(i)},\hat{u}_i^n|f_i,b_i,m_0,f,m^*)$ of \eqref{eqn:ach_shrt_proof} to obtain $(\hat{u}^n_{(i)},\hat{u}_i^n)$ as an estimate of $(u^n,u_i^n)$, for $i=1,2$. Then the processors produce $x^n$ and $y^n$ according to $p(x^n|\hat{u}^n_{(1)},\hat{u}_1^n)$ and $p(y^n|\hat{u}^n_{(2)},\hat{u}_2^n)$, respectively.

To eliminate the extra shared randomness without disturbing the desired i.i.d. distribution on $X,Y$, we need a third set of constraints on rates. Under these constraints below, $(X^n,Y^n)$ and $(F,F_1,F_2)$ are approximately independent~\cite[Theorem 1]{YassaeeAG14}.
\begin{align}
 \hat{R_0}& < H(U|X,Y)\nonumber\\
 \hat{R_1}+\hat{R_0}& < H(U,U_1|X,Y)\nonumber\\
 \hat{R_2}+\hat{R_0}& < H(U,U_2|X,Y)\nonumber\\
 \hat{R_1}+\hat{R_2}+\hat{R_0}& < H(U,U_1,U_2|X,Y)\label{eqn:theorem:arxiv_achievability_gather_3}
 \end{align}
 All these three sets of rate constraints {\color{black}\eqref{eqn:theorem:arxiv_achievability_gather_1}, \eqref{eqn:theorem:arxiv_achievability_gather_2} and \eqref{eqn:theorem:arxiv_achievability_gather_3}} guarantee the existence of a particular realization of random binning (so that we can replace $P$ with $p$ in \eqref{eqn:Thm_ach_protB} and denote the resulting p.m.f. by $\hat{p}$) such that
\begin{equation*}
\hat{p}(x^n,y^n,f_1,f_2,f)\approx p^{\text{Unif}}(f_1)p^{\text{Unif}}(f_2)p^{\text{Unif}}(f)p(x^n,y^n), 
\end{equation*}
which further implies that there exists instances $f^*, f_1^*, f_2^*$ of $F,F_1,F_2$ such that
\begin{equation}
\hat{p}(x^n,y^n|f_1^*,f_2^*,f^*)\approx p(x^n,y^n).
\end{equation}
Note that above equation is the required correctness condition. Elimination of extra shared randomness rates $\hat{R}_0,\hat{R}_1$ and $\hat{R}_2$ from all the rate constraints 
{\color{black}\eqref{eqn:theorem:arxiv_achievability_gather_1}, \eqref{eqn:theorem:arxiv_achievability_gather_2} and \eqref{eqn:theorem:arxiv_achievability_gather_3}} gives us
\begin{align}
R^*&>I(U_1;U_2|U),\nonumber\\
R_0+R^*&>I(X,Y;U),\nonumber\\
R_0+\tilde{R}_1+R^*&>I(X,Y;U,U_1),\nonumber\\
R_0+\tilde{R}_2+R^*&>I(X,Y;U,U_2),\nonumber\\
R_0+\tilde{R}_1+\tilde{R}_2+R^*&>I(U_1;U_2|U)\nonumber\\
&\hspace{12pt}+I(X,Y;U,U_1,U_2)\label{eqn:ach_prefinal}.
\end{align} 
Noting that $R=0.5R_0+R^*$ and $R_i=\tilde{R_i}+0.5R_0$, for $i=1,2$, \eqref{eqn:ach_prefinal} gives $\Rach$.
\end{proof}
\begin{proof}[\textbf{Proof of Theorem~\ref{theorem:unlimited_shared_randomness}}]
For achievability, when rates $R_1, R_2$ are large enough, Theorem~\ref{theorem:achievability} implies that a rate of 
$\max \left\{I(U_1;U_2|U),\frac{1}{2}\big(I(U_1;U_2|U)+I(U;X,Y)\big)\right\}$
is achievable when $X-(U,U_1)-(U,U_2)-Y$. It is easy to see that $U_1=X, U_2=Y$ satisfy the Markov chain $X-(U,U_1)-(U,U_2)-Y$ for any $U$.
So, for any $p(u|x,y)$, if $R=\max\left\{I(X;Y|U),\frac{1}{2}\big(I(X,Y;U)+I(X;Y|U)\big)\right\}$, then there exists $R_1$ and $R_2$ so that $(R,R_1,R_2)\in\mathcal{R}$. Hence, $\RUL\leq \min \max \left\{I(X;Y|U), \frac{1}{2}\big(I(X,Y;U)+I(X;Y|U)\big)\right\}=:R_U$, where the minimum is over all conditional p.m.f.'s $p(u|x,y)$.

For the converse, suppose a rate triplet $(R,R_1,R_2)$ is achievable for $q(x,y)$. Fix an $\epsilon\in(0,\frac{1}{4})$. Then there exists an $(n,2^{nR},2^{nR_1},2^{nR_2})$ simulation code such that 
\begin{equation}
\lVert p_{X^n,Y^n}-q_{X,Y}^{(n)}\rVert_1 < \epsilon\label{eqn:correctness}
\end{equation}
for large enough $n$.
First, we show that there exists a p.m.f. $\gamma_{X,Y,U}$ with $|\mathcal{U}|\leq|\mathcal{X}||\mathcal{Y}|+2$ such that $\lVert \gamma_{X,Y}-q_{X,Y}\rVert_1<\epsilon$ and
\begin{align}
R&\geq I(X;Y|U),\label{eqn1:newthing}\\
R&\geq I(X,Y;U)-g(\epsilon)\label{eqn1:wyner},
\end{align} 
where $\lim_{\epsilon\downarrow 0} g(\epsilon)=0$. We will show \eqref{eqn1:wyner}~\cite{Wyner75}. To obtain \eqref{eqn1:newthing}, we will first show that $nR\geq I(X^n;Y^n|M)$. In Wyner's model~\cite{Wyner75}, the term $I(X^n;Y^n|M)$ is precisely zero. This is not the case here, in general, because of the presence of shared random variables $W_1$ and $W_2$. We will further lower bound the term $I(X^n;Y^n|M)$ by a single-letter form to obtain \eqref{eqn1:newthing}.
\begin{align}
I(X^n;Y^n|M)&\leq I(X^n,W_1;Y^n,W_2|M)\nonumber\\
&=I(W_1;Y^n,W_2|M)\label{eqn1:new_1}\\
&=I(W_1;W_2|M)\label{eqn1:new_2}\\
&\leq I(W_1;M)+I(W_1;W_2|M)\nonumber\\
&\hspace{12pt}-I(W_1;W_2)\label{eqn1:new_3}\\
&=I(W_1;M,W_2)-I(W_1;W_2)\nonumber\\
&=I(W_1;M|W_2)\nonumber\\
&\leq H(M|W_2)\nonumber\\
&\leq H(M)\nonumber\\
&\leq nR\label{eqn1:new},
\end{align}
where \eqref{eqn1:new_1} and \eqref{eqn1:new_2} follow from the Markov chain $X^n-(M,W_1)-(M,W_2)-Y^n$, \eqref{eqn1:new_3} follows because $W_1$ is independent of $W_2$ and $I(W_1;M)\geq0$. 

Let $T$ be a random variable uniformly distributed over $[1:n]$ and independent of all other variables. Then, by continuing \eqref{eqn1:new}, we have
\iftoggle{paper}
{\begin{align}
nR&\geq I(X^n;Y^n|M)\nonumber\\
&=\sum_{i=1}^n I(X_i;Y^n|M,X^{i-1})\nonumber\\
&\geq \sum_{i=1}^n I(X_i;Y_i^n|M,X^{i-1},Y^{i-1})\nonumber\\
&\geq \sum_{i=1}^n I(X_i;Y_i|M,X^{i-1},Y^{i-1})\nonumber\\
&=\sum_{i=1}^n I(X_i;Y_i|U_i)\label{eqn:converse_newthing_1}\\
&=nI(X_T;Y_T|U_T,T),\label{eqn:converse_newthing}
\end{align}}
{\begin{align}
nR&\geq I(X^n;Y^n|M)\nonumber\\
&=\sum_{i=1}^n I(X_i;Y^n|M,X^{i-1})\nonumber\\
&=\sum_{i=1}^n I(X_i;Y^{i-1}|M,X^{i-1})\nonumber\\
&\hspace{2cm}+\sum_{i=1}^n I(X_i;Y_i^n|M,X^{i-1},Y^{i-1})\nonumber\\
&\geq \sum_{i=1}^n I(X_i;Y_i^n|M,X^{i-1},Y^{i-1})\nonumber\\
&= \sum_{i=1}^n I(X_i;Y_i|M,X^{i-1},Y^{i-1})\nonumber\\
&\hspace{2cm}+\sum_{i=1}^n I(X_i;Y_{i+1}^n)|M,X^{i-1},Y^i)\nonumber\\
&\geq \sum_{i=1}^n I(X_i;Y_i|M,X^{i-1},Y^{i-1})\nonumber\\
&=\sum_{i=1}^n I(X_i;Y_i|U_i)\label{eqn:converse_newthing_1}\\
&=nI(X_T;Y_T|U_T,T),\label{eqn:converse_newthing}
\end{align}}
where \eqref{eqn:converse_newthing_1} follows by defining $U_i=(M,X^{i-1},Y^{i-1})$. Following Wyner~\cite{Wyner75}, we lower bound $R$ in another fashion,
\begin{align}
nR &\geq H(M)\nonumber\\
&\geq I(X^n,Y^n;M)\nonumber\\
&=H(X^n,Y^n)-H(X^n,Y^n|M)\nonumber\\
&\geq\sum_{i=1}^n [H(X_i,Y_i)-\epsilon^\prime]-\sum_{i=1}^n H(X_i,Y_i|M,X^{i-1},Y^{i-1})\label{eqn:converse_closeness_of_dist}\\
&\geq\sum_{i=1}^n[I(X_i,Y_i;M,X^{i-1},Y^{i-1})-\epsilon^\prime]\nonumber\\
&=\sum_{i=1}^n[I(X_i,Y_i;U_i)-\epsilon^\prime]\nonumber\\
&= n[I(X_T,Y_T;U_T|T)-\epsilon^\prime]\nonumber\\
&= n[I(X_T,Y_T;U_T,T)-I(X_T,Y_T;T)-\epsilon^\prime]\nonumber\\
&\geq n[I(X_T,Y_T;U_T,T)-\epsilon^\prime-\delta]\label{eqn:T_independent}\\
&=nI(X_T,Y_T;U_T,T)-ng(\epsilon).\label{eqn:converse_wyner}
\end{align}
In \eqref{eqn:converse_closeness_of_dist} and \eqref{eqn:T_independent}, $\epsilon^\prime,\delta\rightarrow 0$ as $\epsilon\rightarrow 0$. We show these steps using \eqref{eqn:correctness} (details are in Appendix). In \eqref{eqn:converse_wyner}, $g(\epsilon):=\epsilon^\prime+\delta$, so $g(\epsilon)\rightarrow 0$ as $\epsilon \rightarrow 0$. Now, we claim that we can find a $\gamma_{X,Y,U}$ such that
\begin{align}
\gamma_{X,Y}&=p_{X_T,Y_T},\label{convex_cover_1}\\
I_{\gamma}(X;Y|U)&=I_p(X_T;Y_T|U_T,T),\label{convex_cover_2}\\
I_{\gamma}(X,Y;U)&=I_p(X_T,Y_T;U_T,T),\label{convex_cover_3}\\
|\mathcal{U}|&\leq|\mathcal{X}||\mathcal{Y}|+2\label{convex_cover_4}.
\end{align}
This directly follows from an application of Convex Cover Method~\cite[Appendix C]{GamalK12} along the same lines as Lemma VI.1 of Cuff~\cite{Cuff13}. 
Note that
\begin{align}
\lVert \gamma_{X,Y}-q_{X,Y}\rVert_1&=\lVert p_{X_T,Y_T}-q_{X,Y}\rVert_1\nonumber\\
&\leq\lVert p_{X^n,Y^n}-q_{X,Y}^{(n)}\rVert_1\label{eqn:cuff}\\
&<\epsilon,\nonumber
\end{align}
where \eqref{eqn:cuff} follows from~\cite[Lemma VI.2]{Cuff13}. Let $\mathcal{S}_\epsilon$, for $\epsilon\geq 0$ be defined as the set of all non-negative rates $R$ such that
\begin{align}
R&\geq I(X;Y|U),\\
R&\geq I(X,Y;U)-g^\prime(\epsilon),
\end{align}
for some p.m.f. $p(x,y,u)$ satisfying \eqref{convex_cover_4} and $\lVert p_{X,Y}-q_{X,Y}\rVert_1\leq\epsilon$ with $g^\prime(\epsilon)=g(\epsilon)$, for $\epsilon>0$ and $g^\prime(0)=0$.
Thus,  for every $\epsilon>0$, it follows from \eqref{eqn:converse_newthing}, \eqref{eqn:converse_wyner} and \eqref{convex_cover_1}-\eqref{eqn:cuff} that, $R\in \mathcal{S}_\epsilon$. Using the continuity of total variation distance and mutual information in the probability simplex, we can show that $\underset{\epsilon>0}{\bigcap}\mathcal{S}_\epsilon=\mathcal{S}_0$ along  the same lines as~Yassaee et al~\cite[Lemma 6]{YassaeeGA15}. Hence $\RUL\geq \min \max \big\{I(X;Y|U),I(X,Y;U)\big\}=:R_L$, where the minimum is over all conditional p.m.f.'s $p(u|x,y)$ with $|\mathcal{U}|\leq |\mathcal{X}||\mathcal{Y}|+2$. So, achievability and converse give us $R_L\leq \RUL\leq R_U$. And since it is trivial to see that $R_L\geq R_U$, we have $\RUL=R_L=R_U$. 
\end{proof}

\begin{proof}[\textbf{Proof of Theorem~\ref{theorem:noshared}}]
For the achievability, it is easy to see from Theorem \ref{theorem:achievability} that $\left(C(X;Y),0,0\right)\in\mathcal{R}$ by identifying that for any $U$ satisfying $X-U-Y$, we have $\left(I(X,Y;U),0,0\right)\in\mathcal{R}$ with the corresponding other auxiliary random variables defined by $U_1=\emptyset$, $U_2=\emptyset$. Hence, $\RNO\leq C(X;Y)$.

 For the converse, suppose $R$ is such that $(R,\epsilon,\epsilon)$ is achievable for every $\epsilon>0$. This implies that for a fixed $\epsilon>0$, there exists an $(n,2^{nR},2^{n\epsilon},2^{n\epsilon})$ simulation code such that  
\begin{align}
\lVert p_{X^n,Y^n}-q_{X,Y}^{(n)} \rVert_1 < \epsilon\label{eqn:arxiv_noshared_1},
\end{align}
for large enough $n$. $R$ can be bounded using \eqref{eqn:arxiv_noshared_1} along the similar lines as \eqref{eqn:converse_wyner}, which gives us 
\begin{align}
R\geq I(X_T,Y_T;U_T,T)-g(\epsilon),\label{eqn:arxiv_noshared_4}
\end{align}
where $\lim_{\epsilon\downarrow 0}g(\epsilon)=0$, $U_T=(M,X^{T-1},Y^{T-1})$ and $T$ is a random variable uniformly distributed over $[1:n]$ and independent of everything else.

 Next, we lower bound $n\epsilon$ in the following fashion.
 \begin{align}
 n\epsilon&\geq H(W_1)\nonumber\\
 &\geq I(W_1;W_2|M)\nonumber\\
 &=I(X^n,W_1;Y^n,W_2|M)\label{eqn:arxiv_noshared_2}\\
 &\geq I(X^n;Y^n|M)\nonumber\\
 &\geq nI(X_T;Y_T|U_T,T),\label{eqn:arxiv_noshared_3}
 \end{align}
 where \eqref{eqn:arxiv_noshared_2} follows from $X^n-(M,W_1)-(M,W_2)-Y^n$, and \eqref{eqn:arxiv_noshared_3} follows along similar lines as \eqref{eqn:converse_newthing}.
 
 Now from \eqref{eqn:arxiv_noshared_4} and \eqref{eqn:arxiv_noshared_3} and using arguments similar to \eqref{convex_cover_1}-\eqref{eqn:cuff} one can show that $R\in\mathcal{M}_\epsilon$, where $\mathcal{M}_\epsilon$ is defined to be the set of all rates $R$ such that
 \begin{align*}
 R&\geq I(X;Y|U)-g(\epsilon),\\
 \epsilon&\geq I(X;Y|U),
 \end{align*}
 where $\lim_{\epsilon\downarrow 0}g(\epsilon)=0$ for some p.m.f. $p(x,y,u)$ satisfying $|\mathcal{U}|\leq|\mathcal{X}||\mathcal{Y}|+2$ and $\lVert p_{X,Y}-q_{X,Y}\rVert_1\leq\epsilon$.
 
 Using the continuity of total variation distance and mutual information in the probability simplex, we can show that $\underset{\epsilon>0}{\bigcap}\mathcal{M}_\epsilon=\mathcal{M}$ along the same lines as~Yassaee et al.~\cite[Lemma~6]{YassaeeGA15}, where $\mathcal{M}$ is defined to be the set of all rates $R$ such that
\begin{align}
R&\geq I(X,Y;U),
\end{align}
for some conditional p.m.f. $p(u|x,y)$ satisfying $X-U-Y$ and $|\mathcal{U}|\leq|\mathcal{X}||\mathcal{Y}|+2$. Hence, $\RNO\geq C(X;Y)$.
\end{proof}

\begin{proof}[\textbf{Proof of Theorem~\ref{theorem:X=Y=Z}}]
For the achievability, we start with non-negative rate triplets $(R,R_1,R_2)$ such that 
\begin{align*}
R+\min\left\{R_1,R_2\right\}&> H(X),\\\nonumber
R&> \frac{H(X)}{2}.
\end{align*}
Now, it is easy to see that such rate triplets are achievable by using Theorem~\ref{theorem:achievability}, i.e., by identifying $U_1=\emptyset, U_2=\emptyset$ and $U$ such that $X-U-X$, it is easy to see that $(R,R_1,R_2)$ is achievable. 

For the converse, suppose a rate triplet $(R,R_1,R_2)$ is achievable for $q(x)$. Then, the existing converse arguments~\cite[Theorem~6]{PrabhakaranS13}, stated in a form suitable to our setting, gives us that there exists $R^\prime\geq0$ and jointly distributed random variables $(Q,U,V)$ such that $U$ \& $V$ are independent and uniformly distributed over $[1:2^{R_1}]$ and $[1:2^{R_2}]$ respectively such that
\begin{align}
R&\geq I(Q;V|U)+R^\prime\label{eqn:arxiv_theorem:X=Y=Z_prabhakaranS_1}\\
R&\geq I(Q;U|V)+R^\prime\label{eqn:arxiv_theorem:X=Y=Z_prabhakaranS_2}\\
H(X)&\leq I(Q;U,V)+R^\prime\label{eqn:arxiv_theorem:X=Y=Z_prabhakaranS_3},
\end{align}
where if $H(X)\geq R_1+R_2$, then we can take $Q=(U,V)$ and if $H(X)\leq R_1+R_2$, then we can take $R^\prime=0$. We have
\begin{align}
R+R_1&\geq I(Q;V|U)+H(U)+R^\prime\label{eqn:arxiv_theorem:X=Y=Z_converse_R+R_1}\\
&\geq I(Q;V|U)+I(Q;U)+R^\prime\nonumber\\
&=I(Q;U,V)+R^\prime\nonumber\\
&\geq H(X)\label{eqn:arxiv_theorem:X=Y=Z_converse_R+R_1_1},
\end{align}
where \eqref{eqn:arxiv_theorem:X=Y=Z_converse_R+R_1} follows from \eqref{eqn:arxiv_theorem:X=Y=Z_prabhakaranS_1} $\&$ $R_1=H(U)$, and \eqref{eqn:arxiv_theorem:X=Y=Z_converse_R+R_1_1} follows from \eqref{eqn:arxiv_theorem:X=Y=Z_prabhakaranS_3}. Similarly, we have 
\begin{align} 
R+R_2\geq H(X)\label{eqn:arxiv_theorem:X=Y=Z_converse_R+R_2}.
\end{align}  
Now, in order to show $R\geq \frac{H(X)}{2}$ we consider two cases.

\underline{\emph{Case (i)}}  $(H(X)\geq R_1+R_2):$

We have
\begin{align}
&2R+R_1+R_2\geq 2H(X)\label{eqn:arxiv_theorem:X=Y=Z_converse_case1_1}\\
\Rightarrow &R\geq \frac{H(X)}{2}\label{eqn:arxiv_theorem:X=Y=Z_converse_case1_2},
\end{align}
where \eqref{eqn:arxiv_theorem:X=Y=Z_converse_case1_1} follows by adding \eqref{eqn:arxiv_theorem:X=Y=Z_converse_R+R_1_1} $\&$ \eqref{eqn:arxiv_theorem:X=Y=Z_converse_R+R_2}, \eqref{eqn:arxiv_theorem:X=Y=Z_converse_case1_2} follows since $H(X)\geq R_1+R_2$.

\underline{\emph{Case (ii)}}  $(H(X)\leq R_1+R_2):$

Notice that when $H(X)\leq R_1+R_2$, we can take $R^\prime=0$ in \eqref{eqn:arxiv_theorem:X=Y=Z_prabhakaranS_1}-\eqref{eqn:arxiv_theorem:X=Y=Z_prabhakaranS_3}~\cite{PrabhakaranS13}. We have
\begin{align}
2R&\geq I(Q;V|U)+I(Q;U|V)\label{eqn:arxiv_theorem:X=Y=Z_converse_case2_4}\\
&= I(Q;V|U)+I(Q,V;U)-I(U;V)\nonumber\\
&= I(Q;V|U)+I(Q,V;U)\label{eqn:arxiv_theorem:X=Y=Z_converse_case2_5}\\
&\geq I(Q;V|U)+I(Q;U)\nonumber\\
&= I(Q;U,V)\nonumber\\
&\geq H(X)\label{eqn:arxiv_theorem:X=Y=Z_converse_case2_6},
\end{align}
where \eqref{eqn:arxiv_theorem:X=Y=Z_converse_case2_4} follows by adding \eqref{eqn:arxiv_theorem:X=Y=Z_prabhakaranS_1} $\&$ \eqref{eqn:arxiv_theorem:X=Y=Z_prabhakaranS_2} with $R^\prime=0$, \eqref{eqn:arxiv_theorem:X=Y=Z_converse_case2_5} follows since $U$ and $V$ are independent random variables, \eqref{eqn:arxiv_theorem:X=Y=Z_converse_case2_6} follows from \eqref{eqn:arxiv_theorem:X=Y=Z_prabhakaranS_3} with $R^\prime=0$.
\end{proof}

\iftoggle{paper}
{\printbibliography}
{
\printbibliography
}
\begin{appendix}
\textbf{Explanation for \eqref{eqn:converse_closeness_of_dist}}.
\balance
\begin{align}
&H(X^n,Y^n)-H(X^n,Y^n|M)\nonumber\\
&=H_p(X^n,Y^n)-H_p(X^n,Y^n|M)\nonumber\\
&\geq H_q(X^n,Y^n)-n\epsilon_1-H_p(X^n,Y^n|M)\label{eqn:arxiv_converse_closeness_of_dist_1}\\
&=\sum_{i=1}^n [H_q(X_i,Y_i)-\epsilon_1]-\sum_{i=1}^n H_p(X_i,Y_i|M,X^{i-1},Y^{i-1})\nonumber\\
\linebreak 
&\geq\sum_{i=1}^n [H_p(X_i,Y_i)-\epsilon_1-\epsilon_2]\nonumber\\
&\hspace{2cm}-\sum_{i=1}^n H_p(X_i,Y_i|M,X^{i-1},Y^{i-1})\label{eqn:arxiv_converse_closeness_of_dist_2}\\
&=\sum_{i=1}^n [H(X_i,Y_i)-\epsilon_1-\epsilon_2]-\sum_{i=1}^n H(X_i,Y_i|M,X^{i-1},Y^{i-1})\nonumber\\
&=\sum_{i=1}^n [H(X_i,Y_i)-\epsilon^\prime]-\sum_{i=1}^n H(X_i,Y_i|M,X^{i-1},Y^{i-1})\label{eqn:arxiv_converse_closeness_of_dist_3}.
\end{align}
We used the following fact in \eqref{eqn:arxiv_converse_closeness_of_dist_1} and \eqref{eqn:arxiv_converse_closeness_of_dist_2}: if two random variables $A$ and $A'$ with 
  same support set $\mathcal{A}$  satisfy $||p_{A} -  p_{A'}||_{1} \leq \epsilon \leq 1/4 $, then it follows from \cite[Theorem 17.3.3]{Cover} that $|H(A) - H(A')|\leq \eta \log |\mathcal{A}|$, where $\eta \rightarrow 0$ as $\epsilon \rightarrow 0$. Now \eqref{eqn:correctness} implies \eqref{eqn:arxiv_converse_closeness_of_dist_1}, where $\epsilon_1\rightarrow 0$ as $\epsilon\rightarrow 0$. Also, note that \eqref{eqn:correctness} implies $\lVert p_{X_i,Y_i}-q_{X,Y} \rVert_1\leq \epsilon$, $\forall i\in [1:n]$, which implies \eqref{eqn:arxiv_converse_closeness_of_dist_2}, where $\epsilon_2\rightarrow 0$ as $\epsilon\rightarrow 0$. In \eqref{eqn:arxiv_converse_closeness_of_dist_3}, $\epsilon^\prime:=\epsilon_1+\epsilon_2$.
  
\textbf{Explanation for \eqref{eqn:T_independent}}. 
\begin{align}
I(X_T,Y_T;T)&=H_p(X_T,Y_T)-H_p(X_T,Y_T|T)\nonumber\\
&\leq H_q(X_T,Y_T)+\delta_1-\frac{1}{n}\sum_{i=1}^n H_p(X_i,Y_i|T=i)\label{eqn:arxiv_T_independent_1}\\
&= H_q(X_T,Y_T)+\delta_1-\frac{1}{n}\sum_{i=1}^n H_p(X_i,Y_i)\nonumber\\
&\leq H_q(X_T,Y_T)+\delta_1-\frac{1}{n}\sum_{i=1}^n[H_q(X_i,Y_i)-\delta_2]\label{eqn:arxiv_T_independent_2}\\
&=H_q(X_T,Y_T)-H_q(X_T,Y_T)+\delta_1+\delta_2\nonumber\\
&=H_q(X_T,Y_T)-H_q(X_T,Y_T)+\delta\label{eqn:arxiv_T_independent_3}\\
&\leq \delta\nonumber.
\end{align}
We used the following fact in \eqref{eqn:arxiv_T_independent_1} and \eqref{eqn:arxiv_T_independent_2}: if two random variables $A$ and $A'$ with 
  same support set $\mathcal{A}$  satisfy $||p_{A} -  p_{A'}||_{1} \leq \epsilon \leq 1/4 $, then it follows from \cite[Theorem 17.3.3]{Cover} that $|H(A) - H(A')|\leq \eta \log |\mathcal{A}|$, where $\eta \rightarrow 0$ as $\epsilon \rightarrow 0$. Now using \cite[Lemma VI.2]{Cuff13}, \eqref{eqn:correctness} implies $\lVert p_{X_T,Y_T}-q_{X,Y}\rVert_1\leq\epsilon$, which implies \eqref{eqn:arxiv_T_independent_1} and \eqref{eqn:arxiv_T_independent_2}. In \eqref{eqn:arxiv_T_independent_3}, we defined $\delta:=\delta_1+\delta_2$, where $\delta\rightarrow 0$ as $\epsilon\rightarrow 0$.
\end{appendix}

\end{document}